Anomalous and Planar Hall Effects in $Co_{1-x}Ho_x$ Near Magnetic Sublattice Compensation.

Ramesh C Budhani[1*], Rajeev Nepal[1], Vinay Sharma[1] and Jerzy T. Sadowski[2]

1. Department of Physics, Morgan State University, Baltimore MD 21251.

2. Center for Functional Nanomaterials, Brookhaven National Laboratory, Upton, NY 11973.

* Ramesh.budhani@morgan.edu

ABSTRACT

Metallic amorphous ferrimagnets derived from alloying *3d* transition metals with *4f* – electron rare earths host fascinating effects of compensation between the *3d* and *4f* magnetic sublattices. Here, a detailed study of the anisotropic magnetoresistance ($\Delta\rho_{xx}$), planar Hall effect ($\rho_{xy}^{PHE}$) and anomalous Hall effect ($\rho_{xy}^{AHE}$) are reported on a series of $Co_{1-x}Ho_x$ thin films over a wide field – temperature (H-T) phase space. Close to magnetic compensation temperature, the $\rho_{xy}^{AHE}$ -H loops show a double sign reversal and signatures of spin – flop transition at higher fields. The $\Delta\rho_{xx}$ and $\rho_{xy}^{PHE}$ also display strong deviations from the classical angular dependence seen in soft ferromagnets like permalloy as the angle $\phi$ between in-plane current and magnetic field is scanned from 0 to $2\pi$. It is argued that the non-zero orbital angular momentum of Ho ions in the lattice and stabilization of bubble domains below magnetic saturation may be responsible for such features. Direct imaging of magnetic textures with X-ray photoelectron microscopy shows formation of stripe domain patterns in the regime of sublattice compensation. Such stripes are likely to transform into magnetic bubbles before full saturation is reached in a large magnetic field.

I. INTRODUCTION

The competing interactions between the two magnetic sublattices of a ferrimagnet, which manifest prominently in several electromagnetic responses of the material, if tuned appropriately through chemical composition, spin torques, and stress fields, can be exploited for spintronic applications [1-3]. From the viewpoint of fundamental physics, the phenomenon of sublattice magnetization compensation, where the net magnetization of the material reaches a minimum [4,5], spin-flop transition marked by breaking of the collinearity between the two magnetization vectors [6,7], and the formation of non-trivial spin textures [8-10], have attracted much attention in recent years. Similarly, the manipulation of magnetic states with spin orbit and/or spin transfer torque, which is quite effective in ferrimagnets near the compensation point, has been a topic of considerable interest [2,3,11-16]. The compensated spin state has also been found more receptive to ultrafast switching with femtosecond light pulses [16-19]. The archetypal ferrimagnets that display robust compensation and compensation-driven electromagnetic responses are alloys of *3d* transition metals with rare-earth elements of the partially filled *4f* electronic shells. Amongst these also, the most studied systems are amorphous alloys of Co with Gd, which yield a perfectly collinear sublattice magnetization [4,5, 21,22]. Additionally, to reduce the magneto-crystalline anisotropy associated with Co and to increase the net magnetic moment of the *3d* sublattice, the GdCo alloys where 15 to 20 % Co is replaced by Fe are also well studied [22,23]. The replacement of Gd with other *4f* elements affects the collinearity of *3d* and *4f* sublattice magnetizations. It has been shown that the collinearity deviations increase as heavier *4f* elements are used. Here, while the 3d sublattice is still strongly ferromagnetic, the *4f* moments may be distributed almost randomly with their directions splayed between $\pi/2$ to $\pi$ away from the direction of the *3d* sublattice magnetization. This, so-called Speri magnetic state is quite robust in alloys where the *4f* element is Dy or Ho [6,7,24,25].

Recent studies of the magnetic state in *3d - 4f* amorphous alloys near the compensation temperature ($T_{comp}$) in systems where the *4f* element Gd is replaced by Tb have yielded some fascinating results [26,27]. It is worth noting that the electrons in the *4f* state of Tb have a non-zero orbital angular momentum, which promotes a large magnetic anisotropy and drives a first-order phase transition from collinear to non-collinear spin state near compensation. Further, the magnetic field – temperature (*H-T*) phase diagram of this transition in CoFeTb alloys changes when the anisotropy of *3d* sublattice is enhanced by coupling to a strongly spin-orbit coupled non-magnetic metal like tantalum [12-15].

In the light of these recent studies, it is important to investigate the magnetic state of the *3d – 4f* alloys where the *4f* atom is in a higher angular momentum state. One such system is presented by the amorphous alloys of holmium and cobalt. The *4f* shell of Ho has the highest angular momentum, which leads to a strong magnetic anisotropy. Lorentz electron microscopy of the magnetic domains in Ho-Co films made by thermal evaporation and sputtering revealed a rich stripe domain pattern, which transforms into bubble domains in low magnetic fields [28,29]. A substantive study of anomalous Hall effect (AHE) in Ho-Co films of unknown composition but with a compensation temperature of ~ 330 K has been presented by Ratajezak and Goscinska [30]. The anomalous Hall resistivity ($\rho^{AHE}$) loops of these flash evaporated films of thickness > 85 nm, show a double loop of the type seen in GdCoTb [26,27], but at temperatures above as well as below the $T_{comp}$. The double loop feature of $\rho^{AHE}$(H) is absent in thinner films. The authors [30] have attributed these anomalous loops to nanoscale inhomogeneities in these films, which could be an artifact of the flash evaporation process due to differences in the vapor pressures of Co and Ho. For thinner films, the measurements of $\rho^{AHE}$ as a function of angle between the field and film normal show a diverging anisotropy field at T~ $T_{comp}$.

This background information on the nature of $\rho^{AHE}$(H) loops near the compensation point in amorphous alloys of CoGd, CoFeGd, CoFeTb and CoHo suggests the need to further investigate the phenomena of double hysteresis in compensated *3d - 3f* alloys and address its dependence on the nature of the *4f* element, its anisotropy and its relative concentration with respect to the *3d* element in the alloy. In addition to $\rho^{AHE}$(H,T), the measurements of anisotropic magnetoresistance (AMR) and planar Hall effect (PHE) in thin films of magnetically ordered systems bring forth valuable information about the magnetic anisotropies [31,32], scattering of charge carriers by spin textures [33], and spin-orbit coupling driven band topology [34,35]. However, it is important to note that there is only a scanty information [30] on the in-plane anisotropy of charge transport in *3d - 4f* compensated systems. Here we present a comprehensive study of AHE, magnetoresistance and anisotropic magneto-transport over a wide range of magnetic field and temperature in a series of 25 nm thick $Co_{1-x}Ho_x$ films, made by co-sputtering of Co and Ho targets. We compare the transport parameters of these alloy films with those of elemental Co and Ho. The transport data are augmented by the measurements of temperature dependent magnetization (M(T)) and imaging of magnetic domain patterns with X-ray photoelectron microscopy. The section-wise contents of this paper are as follows: Section II describes the experimental methods used to prepare and characterize the elemental and alloy films. The results of various measurements and their significance are presented in Section III, with subdivisions as; III(a) AHE of elemental films, III(b) the behavior of AHE in $Co_{1-x}Ho_x$ alloy films in the regimes of sublattice compensation and spin flop transition, III(c) the behavior of AMR and PHE in the vicinity of compensation. A summary of the key findings of this research are presented in Section IV.

## II. EXPERIMENTAL DTAILS

Thin films of $Co_{1-x}Ho_x$ were deposited on c-plane sapphire substrates placed on a rotating platform by dc magnetron sputtering in a load-lock all-metal-seal vacuum chamber of base pressure 8 x $10^{-9}$ Torr. Elemental targets of cobalt and holmium, with purity 99.995% were co-sputtered in a confocal geometry at 5 mTorr pressure of semiconductor grade argon (99.9999%). A ≈ 2 nm thick wetting layer of tungsten (W) was deposited on the substrates before commencing the co-deposition of Ho and Co. Films of different Co/Ho ratio were realized by changing the sputtering power of the holmium source from 15 to 30 watts while the Co target was sputtered at a constant power of 160 watts. A 3 nm thick caping layer of W was deposited on top of the 25 nm thick $Co_{1-x}Ho_x$ layer for protection against oxidation. While the elemental concentration of Co and Ho in the films was controlled nominally from the deposition rate and density of each element, X-ray photoelectron spectroscopy (XPS)-based microanalysis of some of the films yielded the correct composition as listed in Table I. The XPS based elemental analysis also allowed to establish a correlation between sputtering power of Ho target and composition of the films. Shadow masks were used during deposition to create 200 x 3000 $\mu m^2$ Hall bar patterns of the alloy film for measurements of longitudinal resistivity ($\rho_{xx}$) and Hall resistivity ($\rho_{xy}$) in the temperature range of 2 to 400 K in a 9-tesla physical property measurement system (PPMS). The use of a vertical sample rotator allowed in-plane and out-of-plane rotation of the film with respect to the direction of the dc magnetic field for measurements of anisotropic magnetoresistance and planar Hall effect. The static magnetization of the samples was measured from 2 to 300 K by using the vibrating sample magnetometer attachment of the PPMS. We have also measured the $\rho_{xx}$ and $\rho_{xy}$ of elemental cobalt and holmium films. While there is a rich literature on electron transport in thin films of cobalt, such measurements on films of the highly anisotropic rare-earth ferromagnet holmium are lacking. Lastly, the magnetic textures in some of the $Co_{1-x}Ho_x$ films have been imaged with spatially resolved X-ray magnetic circular dichroism (XMCD) measurements conducted at the XPEEM/LEEM endstation of the ESM beamline (21-ID) [36] of the National Synchrotron Light Source (NSLS-II) facility of Brookhaven National Laboratory. XMCD imaging was conducted at room temperature at photon energy 781.6 eV, corresponding to Co-$L_3$ absorption edge.

*Table – I: The compensation temperature of five different samples (A through E) of $Co_{1-x}Ho_x$ and the elemental concentration (atomic %) of Co and Ho in samples A, C and E, as determined by X-ray photoelectron spectroscopy. The composition of B and D was set by adjusted the power of Ho target. The star symbol indicated nominal composition. $T_{comp}$ is the temperature (K) at which the Hall resistivity changes sign.*

| Film | A | B* | C | D* | E |
|---|---|---|---|---|---|
| Co | 81.4 | 78 | 75.8 | 72 | 69.9 |
| Ho | 19.6 | 22 | 24.2 | 28 | 30.1 |
| $T_{comp}$ | 120 | 170 | 245 | 303 | 356 |

## III. RESULTS AND DISCUSSION

### III(a) Sign of Anomalous Hall Resistivity in Elemental Holmium and Cobalt Films

The magnitude and sign of the anomalous Hall resistivity of the $3d$ – $4f$ amorphous alloys are a sensitive function of the intrinsic AHE in the ordered state of these elements, the relative orientation of $3d$ and $4f$ magnetic sublattices, and the topological contribution of magnetic textures present in the alloy films. It is pertinent to first discuss the magnitude and sign of AHE in elemental cobalt and holmium. While the AHE in 3d magnetic metals has been addressed extensively [37,38], measurements of Hall effect in holmium are rather scanty [39]. The Néel temperature ($T_N$) of holmium in single crystal form is 132 K, at which the moments in the basal plane of the hexagonal structure order ferromagnetically but the ordering direction rotates in successive ab-planes forming a helical structure. Below 20 K, this material is ferromagnetic. Measurements on single crystals show a negative Hall coefficient at $T < T_N$ for field direction perpendicular to the ab-plane [39,40]. While the $T_N$ drops with film thickness in ploycrystalline films [41], to the best of our knowledge measurements of AHE in thin films of holmium are not reported. We have measured the $\rho_{xx}$ and $\rho_{xy}$ of a 25 nm thick Ho film over a range of temperatures spanning from 2 to 150 K. Figure 1(a) shows the longitudinal resistivity of holmium films measured at 0.3 and 5 tesla field applied perpendicular to the plane of the film. A distinct change in slope of $\rho_{xx}(T)$ plot is seen in the temperature range of 110 to 130 K, which is highlighted in the $d\rho_{xx}(T)/dT$ vs. T plot shown in the inset whose inflexion point yields a $T_N$ of ≈ 120 K. The field dependence of $\rho_{xy}$ at several temperatures between 2 and 80 K is presented in Fig. 1(b). The Hall resistivity at all temperatures is negative and its non-linear field dependence at lower temperatures, where it is also hysteretic (shown in supplementary Fig. S$_1$), is indicative of a magnetically ordered material. The inset of Fig. 1(b) shows the $\rho_{xy}$ (T) of this film measured at 0.3 and 5 tesla. The anomalous Hall resistivity of Ho shows a strong field dependence, and its sign is negative.

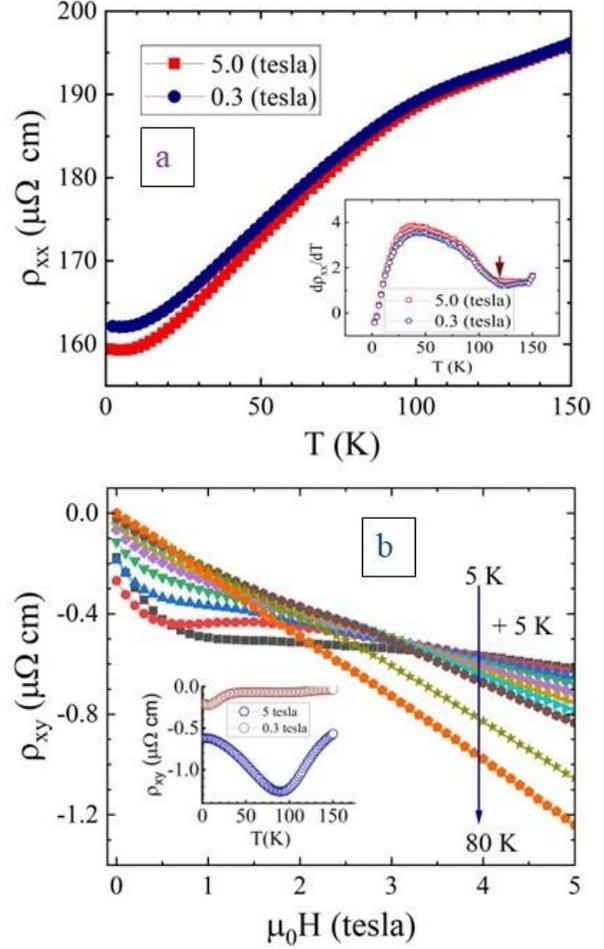

*Fig. 1: (a) Longitudinal resistivity of a 25 nm thick holmium film measured at 0.3 and 5 tesla field applied perpendicular to the film plane. The inset show temperature derivative of the resistivity to highlight the magnetic transition. (b) Hall resistivity of a 25 nm thick holmium film as a function of magnetic field at several temperatures increasing in steps of 5 K, from 5 K to 80 K. The negative sign of $\rho_{xy}$ is noteworthy. Inset shows temperature dependence of $\rho_{xy}$ at 0.3 and 5 tesla.*

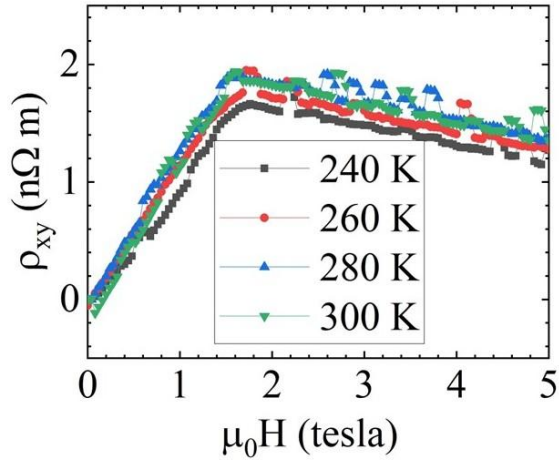

*Fig. 2: Hall resistivity ($\rho_{xy}$) of a 25 nm thick cobalt film measured at 240, 260, 280 and 300 K. The anomalous $\rho_{xy}^{AHE}$ saturates at $\approx$ 1.8 tesla. The behavior of normal Hall resistivity ($\rho_{xy}^{Nrml}$) is represented by the negative slope of the curves beyond saturation indicating a negative Hall coefficient due to the Lorentz force on charge carriers.*

For the sake of completeness, the AHE measurements have been performed on a 25 nm thick polycrystalline film of cobalt as well. Fig. 2 shows the Hall resistivity of this film measured at 240, 260, 280 and 300 K. The anomalous $\rho_{xy}^{AHE}$ is positive and saturates at $\approx$ 1.8 tesla. The behavior of normal Hall effect ($\rho_{xy}^{Nrml}$) is represented by the negative slope of the curves beyond saturation indicating a negative Hall coefficient due to the Lorentz force on charge carriers. The Hall resistivity of elemental cobalt is well documented in the literature [38]. For polycrystalline films of thickness ~ 20 nm, the reported value of $\rho_{xy}$ is 0.22 $\mu\Omega$ cm and 0.08 $\mu\Omega$ cm at 300 and 100 K respectively. Our measurements at room temperature show a saturation value of 0.18 $\mu\Omega$ cm. The field-dependence, sign and numerical values of $\rho_{xy}^{Nrml}$ and $\rho_{xy}^{AHE}$ are consistent with the published results [38]. These measurements on elemental films established two facts which will be useful in understanding the AHE in $Co_{1-x}Ho_x$ alloys. These are: first, the AHE of Ho is negative while its sign in Co is positive, and second, the $\rho_{xy}^{AHE}$ of Ho is larger than that of cobalt in the magnetically ordered state.

### III(b) Magnetic Sublattice Compensation and Spin-Flop Transition in $Co_{1-x}Ho_x$ Films

The anomalous Hall resistivity (AHR) of *3d - 4f* compensated ferrimagnets reaches a minimum value at the sublattice compensation temperature [12,13,26,42]. To illustrate this point for $Co_{1-x}Ho_x$, we show in Fig. 3 the full $\rho_{xy}$(H) loops of the sample with the lowest Ho concentration (film A of Table I), at four temperatures across the compensation temperature. The out-of-plane magnetic field in these measurements was scanned from 0 to + $H_{max}$ to – $H_{max}$ to +$H_{max}$. Fig. 3(a) shows the loop at 110 K. The red and blue arrows in the figure mark the trace of $\rho_{xy}$ (H) as the field is scanned from + 9 to – 9 tesla and then in the reverse direction, respectively. The virgin branch of the $\rho_{xy}$(H) loop is unmarked. At 110 K, the Hall resistivity at high field ($\approx$ 9 tesla) is negative. On lowering the field to ~ 3.6 tesla, the $\rho_{xy}$ decreases following a dependence of the type $\rho_{xy} \sim (\mu_0 H)^n$ with n $\approx$ 0.4 and then stays constant. The critical fields over which the $\rho_{xy}$ remains constant are marked as $h_1$ and $h_2$ in the figure. Subsequently, the $\rho_{xy}$ drops precipitously on field reversal. This drop is followed immediately by a sharp rise to a positive value at $\mu_0 H \approx -0.8$ tesla. The behavior of $\rho_{xy}$ on further increasing the field in the reverse direction is the same as in the positive field quadrant.

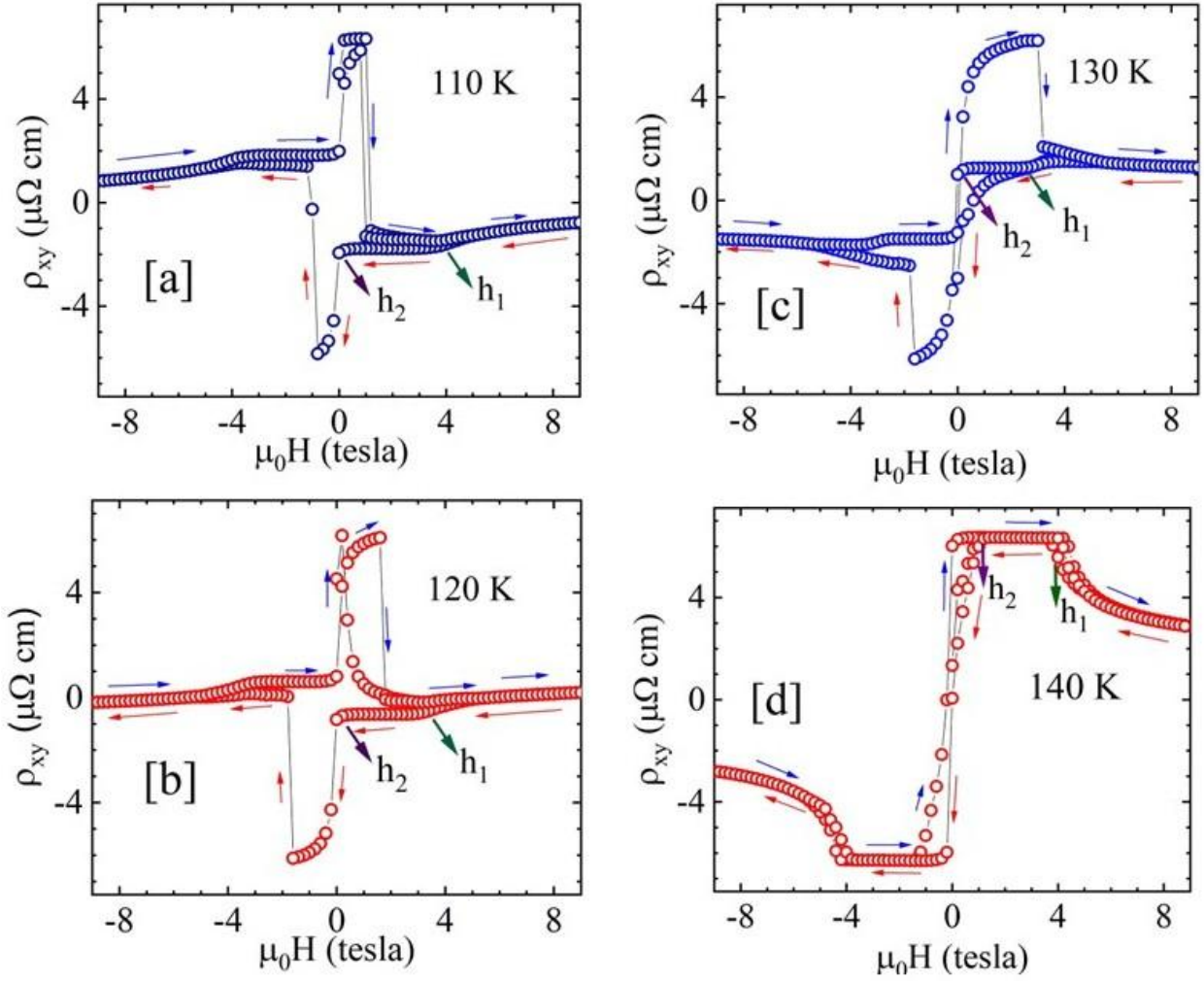

*Fig. 3: Hall resistivity $\rho_{xy}(H)$ of sample A ($Co_{81.4}Ho_{19.6}$) plotted as a function of field at four different temperatures; (a) 110 K, (b) 120 K, (c) 130 K and (d) 140 K across the magnetic compensation temperature. Red and blue arrows in the figures mark the field scans from +9 to -9 and from -9 to +9 tesla respectively. The critical fields $h_1$ and $h_2$ marked in the figures are described in the text.*

For the measurement performed at 120 K, the Hall resistivity at 9 tesla, shown in Fig. 3(b), reverses sign to a positive value, which is smaller by a factor of $\approx 12$ compared to the value of $\rho_{xy}$ at 110 K for the same field. As the field is decreased from 9 tesla, the switchover from a power law to a field-independent $\rho_{xy}$ in this case occurs at a lower field ($\approx 2.8$ tesla). The reversal of field leads to a behavior like that seen at 110 K, including the sharp (a spring-like) rise of $\rho_{xy}$ following the initial drop seen close to zero field. The sign of Hall resistivity at -9 tesla is now negative.

A comparison of the $\rho_{xy}$ at 110 and 120 K places the compensation temperature of the *3d* and *4f* magnetic sublattices of this $Co_{81.4}Ho_{19.6}$ alloy film between these two temperatures. Fig 3(c) shows the loop of $\rho_{xy}$ at 130 K. The Hall resistivity is now prominently positive at + 9 tesla and shows a small but distinct rise ($\rho_{xy} \sim \mu_o H^m$, m ≈ -0.26) till a plateau is reached at ~ 2.2 tesla. A field reversal in this case also leads to a precipitous drop followed by a spring - like rise on increasing the field in the negative quadrant. The sharp features of the Hall loops seen in Fig. 3(a, b and c) change to a much smoother field dependence of $\rho_{xy}$ on increasing the temperature to 140 K (Fig. 3d), but the critical fields $h_1$ and $h_2$ are still definable. Further, there is a distinct difference in the Hall loops at T ≤ 130 K and T ≥ 140 K. This relates to the behavior of $\rho_{xy}$ on bringing the field from 9 tesla to zero. In the former case, the $\rho_{xy}$ remains constant below $h_1$ and then drops in the negative field branch. However, in the latter case it rises rapidly to reach a constant value and then drops on crossing the zero-field line.

The sudden drop in the Hall resistivity on increasing the magnitude of the field in the second and fourth quadrants of the $\rho_{xy}$ (H) loop seen in Fig. 3(d) has been attributed earlier to a spin – flop transition [27]. We track the behavior of the critical fields $h_1$ and $h_2$ at T ≥ 140 K by measuring $\rho_{xy}$ (H) at several temperatures. Since the $\rho_{xy}$ (H) loop at 140 K is fully antisymmetric across the zero-field line, these measurements were performed by monitoring the transverse voltages $V^+_{xy}$ and $V^-_{xy}$ on scanning the field from 0 to +H and then from 0 to -H respectively. The true Hall voltage was obtained by antisymmetrizing the $V^+_{xy}$ and $V^-_{xy}$. The result of these measurements is shown in Fig. 4(a). Clearly, the spin flop transition field $h_1$ increases monotonically on raising the temperature beyond 140 K. The nature of $\rho_{xy}$ (H) between the critical fields $h_1$ and $h_2$ where it remains constant may be understood by considering nature of Hall resistivity in elemental Ho and Co together with the knowledge of how the *3d* and *4f* sublattice magnetizations are oriented in the sample with respect to the external field. The net $\rho_{xy}$ (H) may be expressed as [43],

$$\rho_{xy} = \rho_o H + \mu_0 R_{3d} M_{3d} + \mu_0 R_{4f} M_{4f} \quad --(1)$$

where the first term is the ordinary Hall resistivity resulting from the Lorentz force on free carriers ($\rho_o = 1/ne$, where n is the carrier concentration). The $M_{3d}$ and $M_{4f}$ appearing in the second and third terms are the magnetizations of the *3d* and *4f* metal sublattices with their AHE coefficients $R_{3d}$ and $R_{4f}$ respectively. The free - carrier contribution to Hall is negligibly small in a metallic system. Thus, the magnitude and sign of $\rho_{xy}$ (H) in the present case is dictated primarily by the second and third terms of Eq. 1. From the knowledge of the AHR in elemental Ho and Co presented earlier in Fig. 1(b) and Fig. 2 respectively, it may be concluded that the sign of $\rho_{xy}$ (H) below the compensation temperature, where it is negative, is dictated by the Ho sublattice, whereas above $T_{comp}$, it is the Co sublattice that dominates the $\rho_{xy}$ (H). The fact that the $\rho_{xy}$ (H) remains constant between the critical fields $h_1$ and $h_2$, further suggests that the magnetizations of the two sublattice remain antiparallel. The drop in $\rho_{xy}$ (H) on exceeding $h_1$ indicates a rotation of the cobalt sublattice magnetization away from the direction of the external field. This is expected to happen at the spin – flop transition. In the inset of Fig. 4(a) we show the behavior of critical fields $h_1$ and $h_2$ as a function of temperature. The shaded area in the figure marks the collinear phase which grows on increasing the temperature above the compensation point.

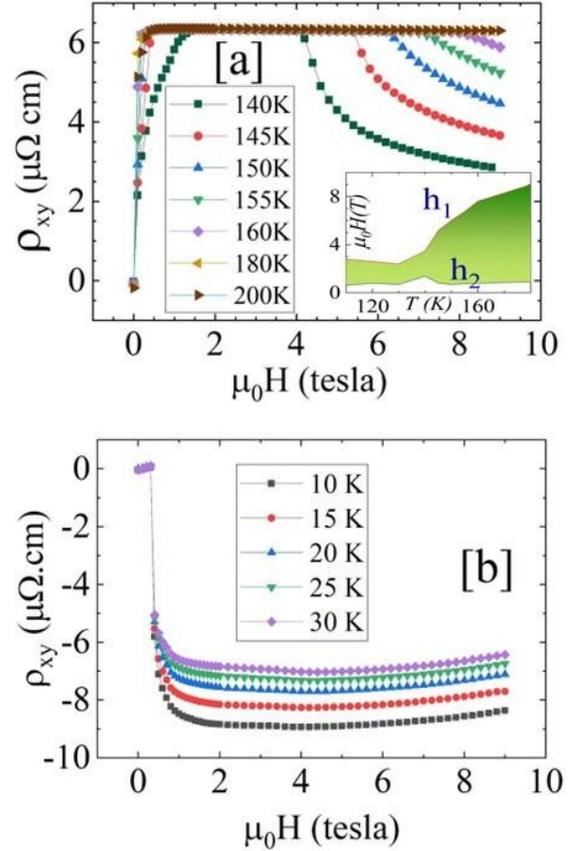

*Fig. 4(a & b): (a) Hall resistivity of sample A at various temperatures (T ≥ 140 K) plotted as a function of increasing field, starting from the true zero-field condition. The sharp drop of $\rho_{xy}$ above a certain field marks the spin reorientation transition in the system. Inset of the figure shows the H-T phase space where $M_{3d}$ and $M_{4f}$ are collinear. (b) The $\rho_{xy}(H)$ of sample A at T < 100 K. The sharp increase of Hall resistivity to a negative value followed by saturation marks the dominant contribution of Ho sublattice to $\rho_{xy}$, which decreases as the temperature is raised.*

The Hall resistivity at lower temperatures, shown in Fig 4(b), has a negative sign and it increases rapidly to reach saturation. A monotonic increase in the saturation value of $\rho_{xy}$ is also seen on lowering the temperature. The sign of $\rho_{xy}$ and increase in its magnitude are suggestive of a dominant contribution of Ho sublattice to $\rho_{xy}$ as seen in Fig. 1(b) for elemental holmium. At lower temperatures, there is no indication of a spin-flop transition ($h_1$ is undefinable), whereas $h_2$ corresponds to field where $\rho_{xy}$ (H) reaches a saturation value.

Based on the results of Fig. 3 and Fig. 4, three regimes of temperature could be defined where the behavior of $\rho_{xy}$ in sample A (Table – I) of CoHo is distinctly different. These are: Regime I for T << $T_{comp}$, regime II for T ≈ $T_{comp}$ and regime III for T > $T_{comp}$. The data corresponding to these three regimes are shown in Fig. 4(b), Fig. 3(b) and Fig. 4(a) respectively. The Hamiltonian presented by Davydova et al. [26,27] that describes the magnetic state of an alloy of *3d* and *4f* electron ions has three relevant terms, the two of which correspond to the magnetic state of the *d* and *f* - electron sublattices and the third term comes from the exchange interaction between the *d* and *f* electron ions. Based on this model, we argue that the behavior in regime I is dominated by the gradually increasing strength of the *4f* sublattice magnetization $M_{4f}$ as the temperature goes well below $T_{comp}$, and by the antiferromagnetic exchange between *3d* and *4f* sublattices, which keeps them collinear. This is consistent with the increasing negative $\rho_{xy}$ as the temperature approached 2 K, together with the absence of any abnormality near zero-field in Fig. 4(b). The behavior of $\rho_{xy}$ in regime II may be less susceptible to exchange as it weakens on increasing the temperature. This seemingly allows the high coercivity of the *4f*-sublattice to dominate. The latter has its origin in the non-zero orbital angular momentum of the Ho atom due to its electronic configuration [Xe]$4f^{11}6s^2$. In earlier studies of *3d - 4f* amorphous alloys [27], the spin flop field $H_{sf}$, has been expressed as $H_{sf} \sim \lambda_{f\text{-}d} (M_{4f} - M_{3d})$, where $\lambda_{f\text{-}d}$ is the sublattice exchange parameter. While this expression assumes a weak magnetic anisotropy of the sublattices, which may be correct for the Gd-based alloys because the orbital angular momentum of *4f* electrons in Gd is zero as the 4f shell is half full, it is not the case for holmium. The Hund's rule for *4f* shell of Ho atom yield the spin and angular momenta quantum numbers as S = 3/2 and L = 6 respectively. Therefore, a strong magnetic anisotropy of the Ho sublattice cannot be ignored. Proof of this comes from the high magnetic coercivity of CoHo alloy films in the temperature regime where the Ho sublattice dominates the magnetization. As seen in Fig. 5, the $H_c$ of a film of composition $Co_{78}Ho_{22}$ rises on lowering the temperature below the compensation point $T_{comp}$. The coercive field has been extracted from the M(H) loops shown in Fig. S3. It is also worth comparing the $H_c$ of the CoHo alloys with the $H_c$ of CoGd films [42]. In the latter case the $H_c$ drops precipitously below the $T_{comp}$ due to the L = 0 state of Gd *4f* shell. This is perhaps the reason why no indications of spin flop are seen at T<< $T_{comp}$ where the Ho sublattice dominates and the exchange interaction between the *3d* and *4f* moments is strong. There are two other noteworthy features of the M(H) loops show in Fig. S3. First, the saturation magnetization extracted from these data and shown in Fig. 5, goes through a minimum in the vicinity of compensation temperature. Secondly, a new critical field $H_{rev}$ above which the magnetization becomes reversible can be defined for these loops. The $H_{rev}$ goes through a peak at the compensation temperature as seen in the inset of Fig. 5. This feature appears to be unique to the CoHo system as it has not been seen in the case of GdCo and GdFeCo films.

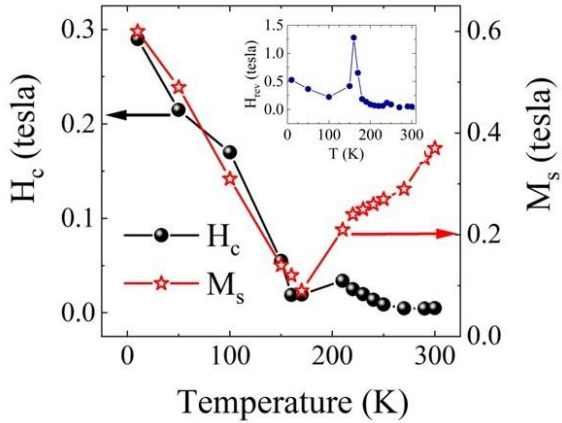

*Fig. 5: Saturation magnetization ($M_s$), coercive field ($H_c$) of sample B ($Co_{78}Ho_{22}$) deduced from dc magnetization loops are plotted as a function of temperature. Inset of the figure shows the field above which magnetization becomes reversible ($H_{rev}$).*

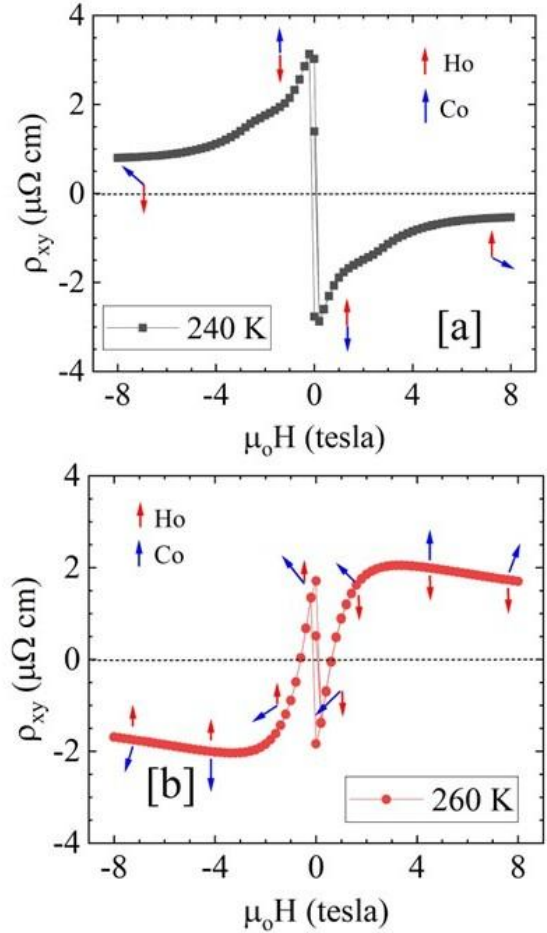

*Fig. 6(a & b): (a) $\rho_{xy}(H)$ loop of sample C at 240 K. The Hall resistivity remains negative in the positive field quadrant of the loop. The red and blue arrows in the figure suggest the orientation of Ho and Co sublattice magnetizations respectively. (b) Shows the Hall resistivity loop at 260 K. Here the $\rho_{xy}$ is positive at higher positive fields and then takes a negative swing at ≈ + 0.8 tesla. The red and blue arrows indicate the likely orientation of Ho and Co sublattice magnetizations respectively to account for the sign of $\rho_{xy}^{AHE}$.*

This dominant role of Ho sublattice diminishes in systems where the compensation temperature is much higher than the $T_N$ of holmium. Here, a new phenomenon seems to appear in the vicinity of $T_{comp}$. This point is best illustrated by the $\rho_{xy}$ (H) data of the sample of composition $Co_{76}Ho_{24}$ (Sample C in Table I) at 240 and 260 K, as shown in Fig. 6(a) and 6(b) respectively. At 240 K and 9 tesla, the $\rho_{xy}$ is small ($\approx$ 0.5 $\mu\Omega$ cm) and negative. Its magnitude increases as the field is brought to zero. On reversing the field, a change in the sign of the $\rho_{xy}$ follows but its magnitude remains the same. A further increase in the field in the reverse direction results in a behavior same as in the first quadrant of the loop. The negative value of $\rho_{xy}$ at 9 tesla suggests that the magnetization of the Ho sublattice ($M_{4f}$) is parallel to the external field. However, its small magnitude indicates that the magnetization of cobalt sublattice ($M_{3d}$) is not antiparallel to $M_{4f}$ but canted at an angle $\Theta < 180$, because a fully antiparallel sublattice magnetizations would yield a larger $\rho_{xy}$. The increase in the magnitude of $\rho_{xy}$ as the field is brought to zero suggests continuous rotation of $M_{3d}$ till $\Theta = 180$, at which point the $\rho_{xy}$ has maximum negative value. The rotation of the $M_{3d}$ and $M_{4f}$ is represented by blue and red arrows respectively in Fig. 6(a), as the field is scanned from +9 to -9 tesla.

However, a similar interpretation for the data taken at 260 K (Fig. 6(b)) where the $\rho_{xy}$ at 9 tesla is positive but changes sign on reducing the field to $\approx$ 0.5 tesla, demands that either the $M_{3d}$ rotates to $\Theta < 90$ or the $M_{4f}$ flips and becomes parallel to external field to account for this change in sign of $\rho_{xy}$. In so much so that such a rotation of $M_{4f}$ or $M_{3d}$ appears unphysical, an alternative explanation is plausible for the data of Fig. 6(b). Since the sign reversal of $\rho_{xy}$ on lowering the field in the first quadrant of the loop disappears when the temperature is raised above $\approx$ 280 K, it could be argued that the anomaly seen in Fig. 6(b) emerges from a topological contribution to Hall resistivity made by some non-trivial spin textures existing in the film prior to reaching a fully saturated magnetic state. Such topological spin textures have been seen in ferrimagnets near the compensation temperature [8-10,44,45]. However, the fact that this anomaly is not observed in the case of the sample with lower Ho concentration, the appearance of spin textures may be concentration dependent.

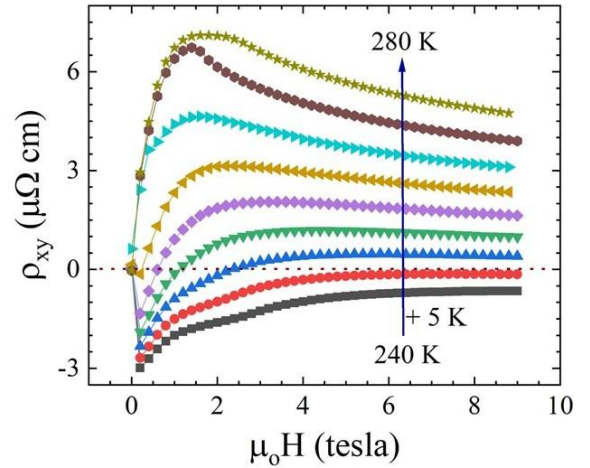

*Fig. 7: Hall resistivity of sample C at various temperatures, increasing from 240 to 280 K in steps of 5 K, plotted as a function of increasing field, starting from the true zero-field condition. Noticeably, the Hall resistivity at $T \leq 245$ K is negative over the entire field range.*

The Hall resistivity of sample C in the field range of 0 to 9 tesla at several temperatures from 240 and 280 K obtained after antisymmetrization of the 0 to +9 and 0 to -9 tesla measurements is shown in Fig. 7. Three features of this figure are noteworthy; first, while the Hall data at T > 265 K is positive over the entire field range, it reaches a peak value and then shows a gradual drop as the maximum field of 9 tesla is approached. Second, at temperatures 250 K ≤ T ≤ 265 K the $\rho_{xy}$ crosses from a negative to positive value as the field approaches 9 tesla. The critical field at which this change in sign occurs increases on lowering the temperature. Thirdly, the $\rho_{xy}$ at T < 250 K remains negative over the entire field range although its magnitude decreases on increasing the field. While the behavior of $\rho_{xy}$ in the temperature regimes I and III, where no change in sign takes place may be attributed to a gradual rotation of $M_{4f}$ and $M_{3d}$ such that the canting angle shifts from 180 degree to a lower value on increasing the field, this may not be true in the temperature regime II. As mentioned earlier in the context of describing Fig. 6b, this behavior is suggestive of a topological contribution to Hall resistivity. A hallmark of systems that show skyrmion or bubble-like spin textures before reaching a fully saturated state is the observation of a stripe or labyrinth type domain pattern in zero field [9,10, 28,29]. Such patterns are clearly seen in the X-ray photoelectron microscopy images of domain patterns in a HoCo film (sample D of Table I) of compensation temperature close to 300 K (Fig. 8). The magnetic contrast in these images comes from the magnetic circular dichroism of Co atoms. The topological features in the Hall resistivity of sample D are presented in Fig. $S_4$. However, here it occurs over a very narrow range of temperature between 304 and 308 K.

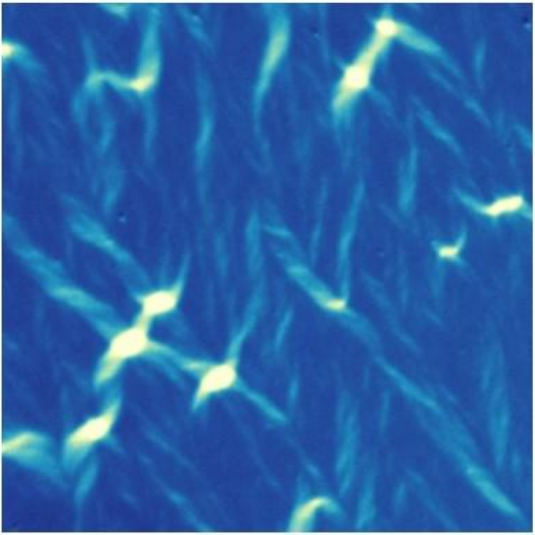

*Fig. 8: X-ray magnetic circular dichroism image of the magnetic texture in sample D at 300 K, taken with a photon energy of 781.6 eV, corresponding to Co-$L_3$ absorption edge. Field-of-view is 20 μm.*

III (C) Anisotropic Magnetoresistance and Planar Hall Effect near the Compensation Temperature

The dramatic changes in the Hall resistivity seen in the regime of $T_{comp}$ are also reflected prominently in the measurements of anisotropic magnetoresistance and planar Hall effect in these samples, which are performed in a geometry as sketched in Fig. 9, where a charge current density $J_x$ flows in the x direction, and the induced electric fields $E_{xx}$ and $E_{xy}$ in the x and y directions respectively due to various scattering processes and band structure effects, are measured as the magnetic field ($\mu_0 H$) is rotated in the xy-plane from −90° to 270°. The angle ϕ = 0 corresponds to the situation when H and $J_x$ are parallel. A distinct ϕ-dependence of the induced electric fields $E_{xx}$ and $E_{xy}$ emerges in the material with an ordered magnetic moment M. In general, the current density J can be expressed in terms of E, H and M as [46],

**J** = $\sigma_{xx}^{(0)}$ **E** + $\sigma_{xy}^{(0)}$ **E** x **H** + $\sigma_{xy}^{(A)}$ **E** x **M** + ($\sigma_{xx}^{(0)}$ τ α/c)[(**M.H**)**E** − (**E.H**)**M**]  --- (2)

Where $\sigma_{xx}^{(0)}$, $\sigma_{xy}^{(0)}$, $\sigma_{xy}^{(A)}$, τ, α and c are the zero-field longitudinal conductivity, Hall

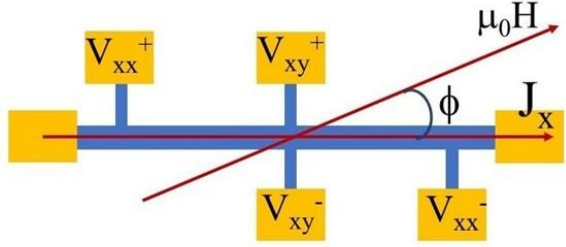

*Fig. 9: A sketch of the AMR and PHE measurement geometry. Sample is placed in the rotating puck of the PPMS in such a way that the current (I) and magnetic field (H) remain in the plane of the sample while the sample is rotated through 0 to $2\pi$.*

conductivity due to the Lorentz force on charge carriers, the anomalous Hall conductivity due to a non-zero magnetization of the sample, carrier scattering time, a numerical coefficient and the speed of light, respectively. For a thin film of a soft magnetic material, placed in an in-plane field where the vectors H and E are coplanar and M follows the direction of H, Eq. 2 yields the anisotropic magnetoresistance contribution to $\rho_{xx}$ and the planar Hall resistivity ($\rho^{PHE}$) as [46-49],

$\rho_{xx} = \rho_\perp - \Delta\rho \cos^2\phi$  ––– 3(a)

$\rho^{PHE} = -\Delta\rho \cos\phi . \sin\phi,$  ––––– 3(b)

where $\Delta\rho$ ( $= \rho_\perp - \rho_{//}$) is defined in terms of longitudinal resistivities $\rho_\perp$ and $\rho_{//}$ when $\phi = 90^0$ and $0^0$ respectively.

Eq.3(a) predicts that the AMR would peak at $\phi = 0$, $\pi$, and $2\pi$ on rotation from zero to 360 degrees if the magnetization M follows the direction of external field. Similarly, the PHE should be optimum at $\pi/4$, $3\pi/4$ and $5\pi/4$. However, deviations from the angular dependence predicted by Eqs. 3(a) and 3(b) may occur due to some trivial as well as some non-trivial causes. The trivial factors are: (i) A non-zero out-of-plane component of the magnetic field resulting from a misalignment of the film plane and the plane of rotation. The anomalous Hall voltage resulting from this misalignment is antisymmetric in field and can be eliminated on symmetrization of the (+H) and (−H) data. (ii) Hall contacts are not on an equipotential line, resulting in a zero – field voltage across the Hall contacts. This signal adds to the PHE voltage on symmetrization of $\rho_{xy}$ [{($\rho_{xy}$(+H) + $\rho_{xy}$(−H)}/2]. However, the misalignment signal can be subtracted from the measured $\rho_{xy}$(H) provided its value is small such that the AMR induced changes in it are insignificant compared to the true $\rho_{xy}$(H). In addition to these two factors, a contaminant to $\rho_{xy}$ comes from the orbital magnetoresistance (OMR) of the misaligned section of the transverse contacts when the sample is tilted with respect to the plane of rotation. This would add a $\cos^2\phi$ term in Eq. (3b). The misalignment of the plane of rotation and the plane of the sample also adds an error in the value of AMR if the OMR of the sample is large.

The non-trivial causes of deviations from the angular dependence predicted by Eq. 3(a) and 3(b) precipitate when the magnetization vector is not parallel to the external field. This issue has been discussed in the literature in the context of AMR in polycrystalline films [49], which brings a polar angle ($\theta$, the angle between M and film normal) dependence as well in Eq. 3(a). The $\theta$-dependence of AMR and PHE is likely to be much more significant in films with perpendicular magnetic anisotropy (PMA). However, there is little prior discussion on AMR and PHE of ferromagnetic films with PMA. To address this issue, we first present in Fig. 10(a) the AMR data of sample C measured at 5 tesla for several temperatures across the $T_{comp}$, which for this sample is between 250 and 260 K. It is clearly seen in the figure that the data taken at T < 220 K follow the dependence predicted by Eq. 3(a). However, large deviations from the $\cos^2\phi$ dependence occur in the temperature window

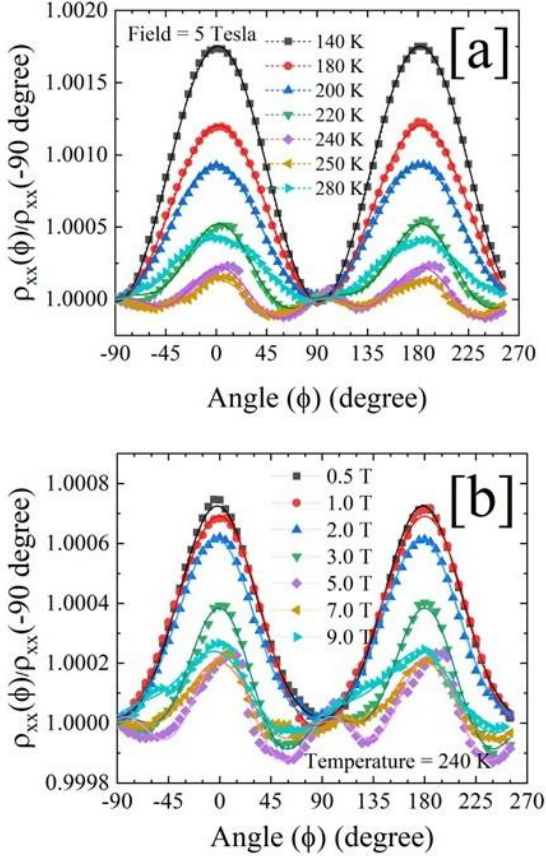

*Fig. 10 (a & b): (a) Anisotropic magnetoresistance of sample C at 5 tesla in the temperature range of 140 to 280 K as the angle ϕ between the current and field is scanned from - 90 to 270 degrees. Anomalous angle dependence of $\rho_{xx}$ is evident in the temperature range of sublattice magnetization compensation. (b) AMR of sample B near the compensation temperature (240 K) at several values of in-plane magnetic field. Here the anomalous behavior is striking at the higher fields (≥ 3 tesla).*

of 220 K < T ≤ 280 K. A distinct feature of these data is the tiny peak in $\rho_{xx}$ at ϕ = 90 degree, which can be modeled by expressing the ϕ dependence of $\rho_{xx}$ as

$\rho_{xx}(\phi) = A + B \sin^2\phi + C \cos^2\phi \sin^2\phi$  —— 4

The $\sin^2\phi$ term becomes dominant in the temperature regime of magnetization compensation due to the out-of-plane component of the magnetization vector, which in the direction of external field is M sinθ. We argued that the 'θ' is a function of the azimuthal angle ϕ due to spin angular momentum transfer from the transport current to the magnetization vector [11]. Here, it is assumed that while the θ varies with ϕ due to the angular momentum transfer from current, the plane formed by the M and H vectors remains unchanged as the sample is rotated from ϕ = 0 to ϕ = 360 degree. It is a fair assumption because the driving field H will always nudge the M in its direction. Further, to account for the even parity of magnetoresistance on field, the first order correction goes as $\sin^2\phi$.

An intriguing aspect of AMR is its dependence on field near the compensation temperature, as shown in Fig. 10(b). Here we note that while the AMR at lower fields (≤ 2 tesla), follows the angular dependence of Eq. 3(a), deviations are seen at the higher fields, which can be expressed by Eq. 4. The important conclusions that one can draw from the data of Fig. 10(a) and 10(b) are that the $\sin^2\phi$ (and hence the polar angle θ) dependence of AMR is relevant only in the field regime where the magnetization is about to saturate, this is also the regime where isolated magnetic textures like bubbles and dipolar skyrmions may appear in the system [8-10, 28,29].

The true PHE (field symmetrized $\rho_{xy}$) at 240 K for several fields and at 5 tesla at different temperatures across the $T_{comp}$ is shown in Fig.11(a) and Fig.11(b) respectively. As in the case of AMR (Fig. 10(a&b)), the ϕ-dependence of PHE also deviated from Eq. 3(b) in the temperature regime close to the compensation temperature and in the field range just below full saturation. The deviation of the PHE from that embodied in Eq 3(b) can be modeled by adding a $\sin^2\phi$ term in Eq.3.

The deviations in PHE and AMR from Eq. 3(a) and 3(b) have been seen in a large

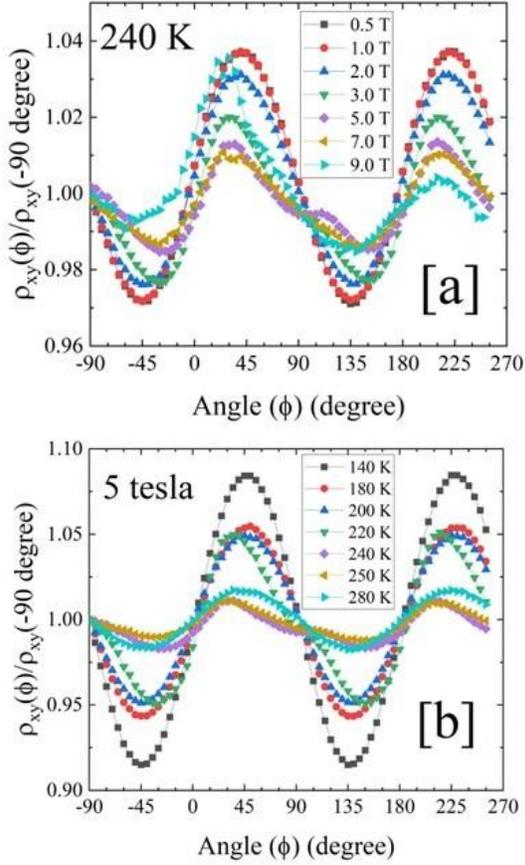

*Fig. 11 (a & b): (a) Planar Hall resistivity of sample C at 240 K for various values on in-plane field. Deviations from the predicted behavior (Eq. 3(b)) are seen at the higher fields. (b) The PHE of sample C measured at 5 tesla at several temperatures across the compensation point.*

number of systems where the magnetization is not parallel to the magnetic field [31] due to inherent magnetocrystalline anisotropy of the magnetic structure or due to extraneous factors such as strain. While the deviations in most cases start from the lowest field, here this anomaly limited to a specific window of temperature and field. This fact strongly suggests that the deviations are tied to some unique magnetic textures that appear in a system with PMA just below magnetic saturation. Clear evidence of this comes from the measurement of PHE in the classic non-centrosymmetric magnet MnSi where the skyrmion phase is stabilized over a narrow temperature range, at a field just below full magnetic saturation [33]. The $\rho^{PHE}$ of MnSi shows a sharp steplike increase when measured as a function of field in the T-H phase space where skyrmions are stabilized. The measurements of AMR in MnSi [33] also show the signatures of a periodically modulated magnetic structure of this system due to skyrmions. In the light of these observations, we suggest that the unique $\sin^2\phi$ dependence of the AMR and PHE, seen in the critical regime of the H-T phase space may emerge due to current induced torque on skyrmions, which are likely to form in these ferrimagnets [8-10].

## IV. SUMMARY

In summary, we have measured anisotropic magneto-transport in thin films of $Co_{1-x}Ho_x$ (x ranging from ≈ 0.19 to 0.30) prepared by cosputtering of Co and Ho targets, over a wide range of temperature and magnetic field. Additionally, the anomalous Hall resistivity of elemental holmium and cobalt films has been measured to understand the sign reversal of $\rho_{xy}^{AHE}$ at the compensation temperature of the alloy films. A vertical rotator allowed in-plane variation of the angle ($\phi$) between the field and transport current from 0 to $2\pi$ to extract $\Delta\rho_{xx}$ ($\phi$) and $\rho_{xy}^{PHE}$ ($\phi$). Similarly, the measurement of $\rho_{xy}$ in the out of plane field orientation yielded the anomalous Hall resistivity loops ($\rho_{xy}^{AHE}$ (H)). While the temperature scans of $\rho_{xy}^{AHE}$ show a sign reversal at the compensation temperature of the *3d* and *4f* magnetic sublattices, the $\rho_{xy}^{AHE}$ loops for a specific composition are characterized by a double sign reversal and the signatures of a spin – flop transition where the *3d* and *4f* sublattices deviate from collinearity. Further, the H-T phase space over which the sublattice magnetizations stay collinear has been deduced. It is also noted that some of the characteristic features of the $\rho_{xy}^{AHE}$ (H) loops are strongly dependent on the relative concentration of Ho and Co. A key result of this study is the observations of strong deviations of $\rho_{xx}$ ($\phi$) and $\rho_{xy}^{PHE}$($\phi$) from the classical $\cos^2\phi$ and $\sin\phi \cos\phi$ dependence respectively, seen in soft ferromagnets. These features are prominent near the compensation temperature. The sign reversal seen in the positive quadrant of the $\rho_{xy}^{AHE}$ (H) loops and anomalous $\rho_{xx}$ ($\phi$) and $\rho_{xy}^{PHE}$($\phi$) are suggestive of a critical role of magnetic textures that are stabilized in compensated ferrimagnets below magnetic saturation. The high coercivity of $\rho_{xy}^{AHE}$ (H) loops below $T_{comp}$ where the Ho sublattice dominates electron transport is indicative of the large orbital angular momentum associate with the Ho ions in this ferrimagnet. Direct imaging of magnetic textures with X-ray photoelectron microscopy shows stripe domain patterns which may have transformed into bubble domains and contributed to the anomalies seen in the angular and field dependencies of $\Delta\rho_{xx}$, $\rho_{xy}^{PHE}$ and $\rho_{xy}^{AHE}$.

## V. ACKNOWLEDGMENTS

This research is funded by the Air Force Office of Scientific Research under Grant No. FA9550-19-1–0082. Partial support has also come from the United States Department of Defense, Grant No. W911NF2120213. The XMCD related work used resources of the Center for Functional Nanomaterials and the National Synchrotron Light Source II, which are U.S. Department of Energy (DOE) Office of Science facilities at Brookhaven National Laboratory, under Contract No. DE-SC0012704. We thank Matthew Kramer and Durga Paudyal at Ames Lab, Iowa for facilitating XPS measurements.

SUPPLIMENTARY INFORMATION

"Anomalous and Planar Hall Effects in $Co_{1-x}Ho_x$ Near Magnetic Sublattice Compensation."

Ramesh C Budhani[1*], Rajeev Nepal[1], Vinay Sharma[1] and Jerzy Sawaski[2].

1. Department of Physics, Morgan State University, Baltimore MD 21251.

2. Center for Functional Nanomaterials, Brookhaven National Laboratory, Upton, NY 11973.

* Ramesh.budhani@morgan.edu

This section describes the results of transport and magnetization measurements performed on thin films of cobalt and holmium prepared under the same conditions as the samples $Co_{1-x}Ho_x$ described in the manuscript. As supporting materials, here we also present the results M(H) loop measurements on sample B across the compensation temperature and Hall resistivity data of sample D whose compensation temperature is ≈ 302 K.

While magneto-transport in thin films of cobalt has been studied extensively [1,2], this is not the case for holmium. This rare-earth element shows a rich magnetic structure. It orders antiferromagnetically at the Néel temperature $T_N$ ≈ 130 K below which the moments order in a helical structure in which the c-axis is the screw axis [3]. Below 20 K, this material is ferromagnetic. The resistivity of thermally evaporated Ho films has been measured by Dudas et al [4]. The $T_N$ of thicker films (198 nm) is marked by a distinct change in the slope of resistivity vs temperature plot at 126 K. For a comparison of the Hall resistivity, we refer to the work of Volkov et al [5] on single crystals of Ho. The $\rho_{xy}$ in the paramagnetic state is negative. Below $T_N$, the $\rho_{xy}$ is anisotropic with anomalous field dependence. The $\rho_{xy}$ vs field plot of the 25 nm film whose $\rho_{xx}$ is shown in Fig. 1(a) is presented in Fig. $S_1$. At 2 K the $\rho_{xy}$ shows a distinct irreversibility below ≈ 3 tesla. However, at 80 K while the irreversibility disappears, the $\rho_{xy}$ still shows some vestige of non-linearity. The Hall resistivity in the paramagnetic state is linear in field, as expected, and the sign of Hall coefficient is negative.

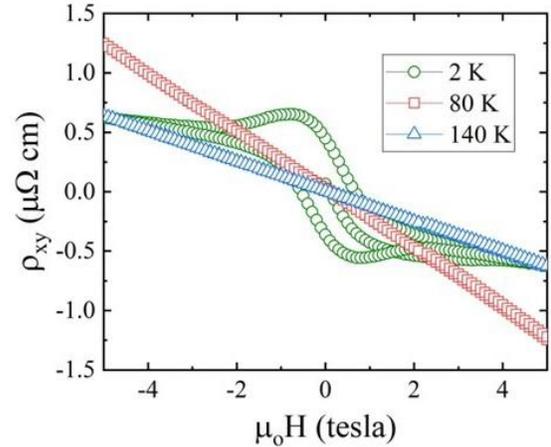

*Fig. $S_1$: The Hall resistivity of the 25 nm thick Ho film measured as a function of magnetic field.*

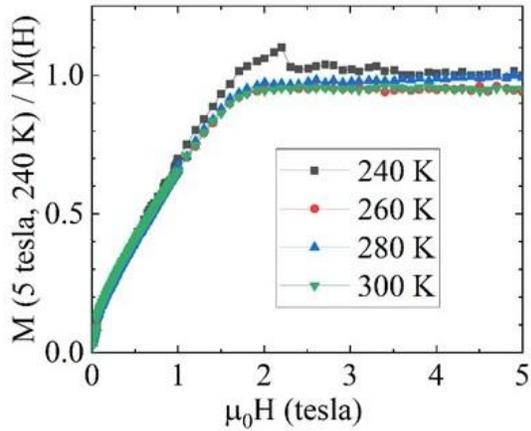

*Fig. S$_2$: Magnetization of a 25 nm thick cobalt film measured as a function of magnetic field directed perpendicular to the plane of the film.*

Figure S$_2$ shows the behavior of magnetization of the 25 nm thick Co films measured in the out-of-plane field geometry. The magnetization of the film reaches saturation at the same field at which the $\rho_{xy}^{AHE}$ saturates, as seen in Fig. 2 of the paper. Clearly, there is an excellent correlation between the anomalous Hall resistivity and magnetization of the sample as predicted by Eq. 1 in the main text.

To establish the compensation temperature of sample B (nominal composition – Co$_{78}$Ho$_{22}$), we have measured the dc magnetization of this sample in the out-of-plane field geometry over a wide temperature range spanning from 10 to 300 K. The M(H) loops are shown in Fig. S$_3$. Below 100 K, the M(H) loops are characterized by a large coercivity, which decreases on increasing the temperature. The saturation magnetization is minimum near 170 K, which marks the compensation temperature of this alloy film. As we go above this temperature, magnetization increases gradually and so does the critical field at which magnetization saturates. This is indicative of a slow tilt of the magnetization vector from the out of film plane to in plane configuration. While this behavior is seen in Gd based compensated ferrimagnets as well [6], the large coercivity at lower temperatures is characteristic of this Ho based alloy, indicating the role of large orbital angular momentum carried by 4f shell of holmium. The other distinctive feature of the M(H) loops is the behavior of the critical field above the magnetization becomes reversible (H$_{rev}$). The H$_{rev}$, goes through a peak near the compensation temperature (see, Fig. 5, main text).

The Hall loops of the sample with nominal composition Co$_{72}$Ho$_{28}$ are shown in Figure S$_4$. The $\rho_{xy}$ (H) shows a dramatic change in the narrow temperature window of 300 to 302 K. It reverses sign at the higher field but at 302 K the Hall resistivity at fields below ≈ 0.5 tesla is still negative. A similar behavior was seen for sample C (Fig. 6b and Fig. 7), which has been attributed to topological Hall effect.

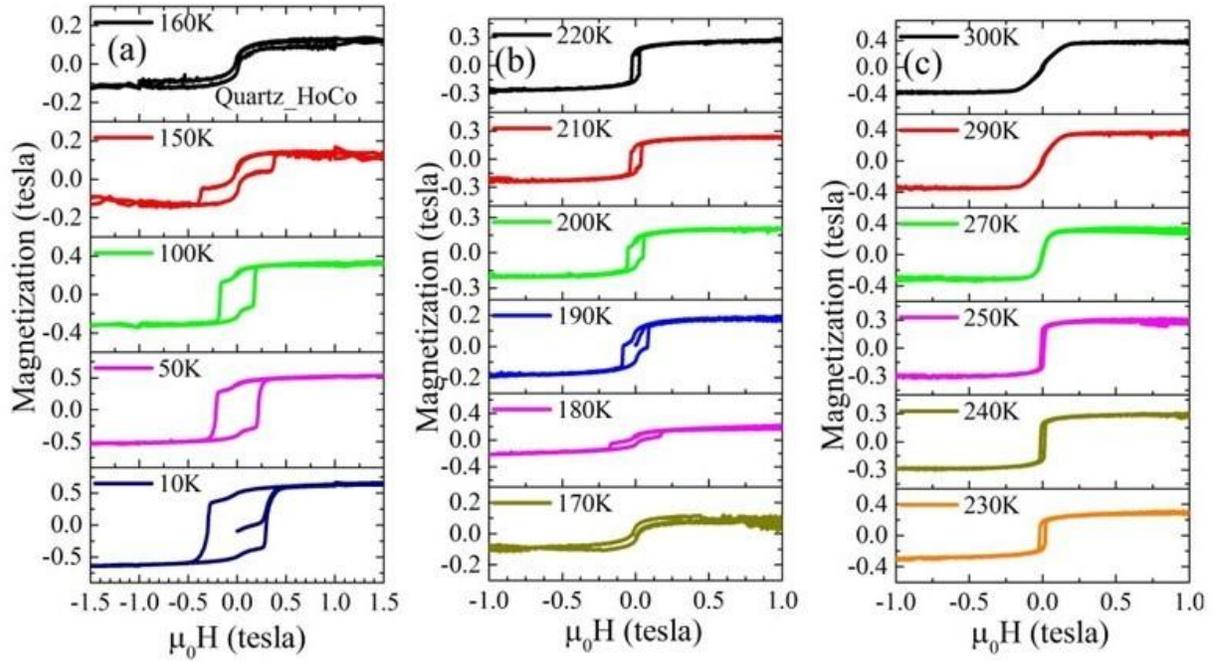

*Fig. S₃: Magnetization vs field loops of the HoCo film (film B of Table I) at several temperatures between 10 and 300 K. The saturation magnetization is minimum at ≈ 170 K, which marks the compensation temperature.*

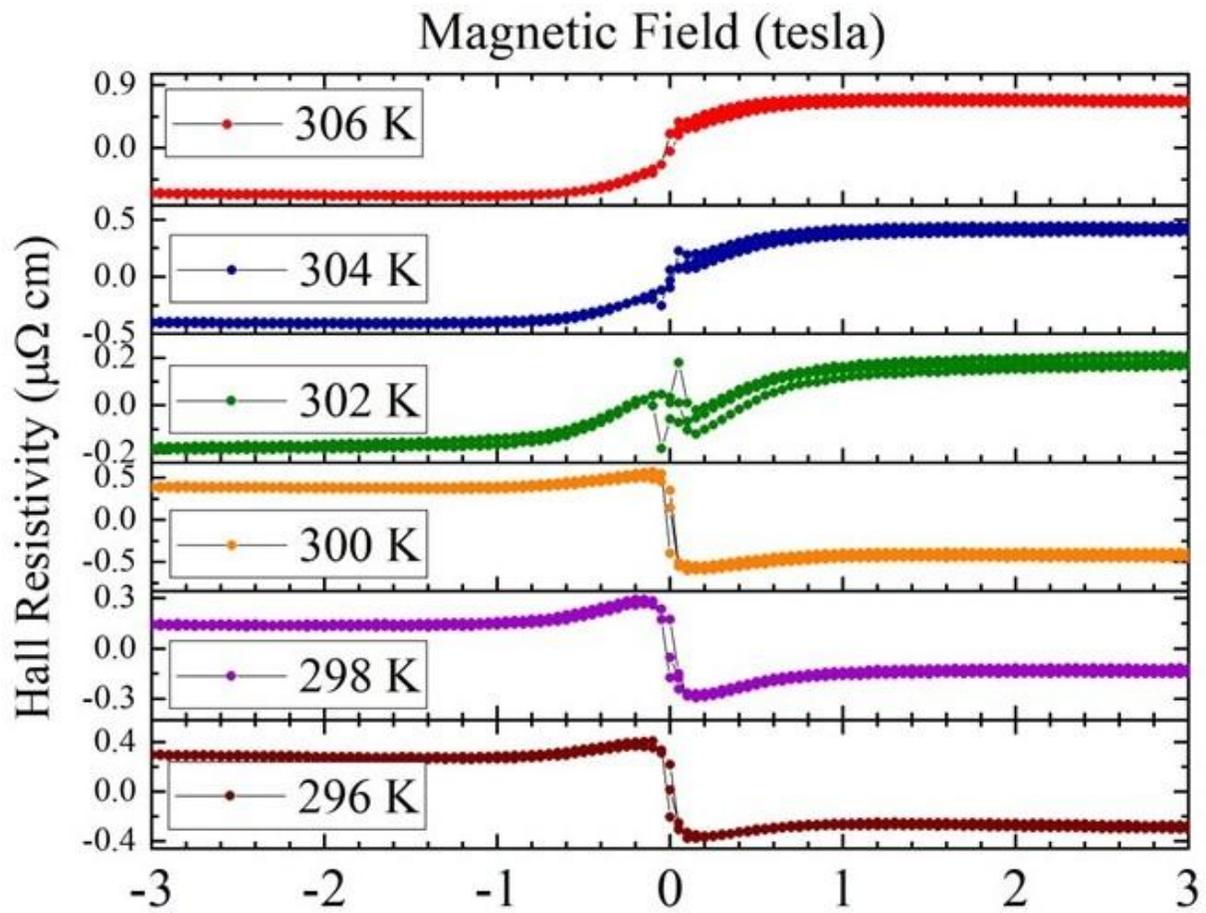

*Fig. S4: Hall resistivity vs field plots of sample D. The change of sign of $\rho_{xy}$ marks the compensation temperature.*